\documentclass[10pt]{iopart}

\usepackage{graphicx}
\usepackage{color}
\usepackage{cite}

\begin{document}

\title[Hybrid SCADW and LLDW treatment of electron-impact single ionisation of W$^{15+}$ and W$^{16+}$]{Hybrid subconfiguration-average and level-to-level distorted-wave treatment of electron-impact single ionisation of W$^{15+}$ and W$^{16+}$}

\author{F~Jin$^{1,2}$,  A~Borovik, Jr.$^2$, B~Ebinger$^{2,3}$ and S~Schippers$^2$}
\address{$^1$Department of Physics, National University of Defense Technology, Fuyuan Road 1, 410022 Changsha, People's Republic of China}
\address{$^2$I. Physikalisches Institut, Justus-Liebig-Universit\"at Gie\ss{}en, Heinrich-Buff-Ring 16, 35392 Giessen, Germany}
\address{$^3$GSI Helmholtzzentrum f\"{u}r Schwerionenforschung GmbH, Planckstr. 1, 64291 Darmstadt, Germany}
\ead{ftjin@nudt.edu.cn}
\vspace{10pt}
\begin{indented}
\item[]\today
\end{indented}

\begin{abstract}
Recently, we have demonstrated (Jin \etal 2020 \textit{J.\ Phys.\ B: At. Mol. Opt. Phys.} \textbf{53} 075201) that a hybrid subconfiguration-average and level-to-level distorted wave treatment of electron-impact single ionisation (EISI) of W$^{14+}$ ions represents an accurate and manageable approach for the calculation of EISI cross sections of a complex ion. Here we demonstrate the more general validity of this approach by comparing hybrid cross sections for EISI of W$^{15+}$ and W$^{16+}$ with the recent experimental results of Schury et al. (2020 \textit{J.\ Phys.\ B: At. Mol. Opt. Phys.} \textbf{53} 015201). Our calculations also account for the resonant-excitation double autoionisation (REDA) process which is important in the electron energy range 370--600 eV and for the possible presence of initially metastable ions in the experiment.
\end{abstract}

%
%
\submitto{\JPB}
%
%
\ioptwocol

\section{\label{intro}Introduction}

Atomic processes involving tungsten ions are of current interest because of the use of tungsten as a wall material in fusion devices such as the ITER tokamak \cite{Skinner2009,Mueller2015b}. One of the processes that govern the charge balance of tungsten in a hot thin plasma is electron-impact single ionisation (EISI). Consequently, EISI of tungsten ions has been repeatedly studied by experiment \cite{Montague1984,Stenke1995c,Rausch2011,Borovik2016}. The most recently published experimental data are those of Schury \etal \cite{Schury2020} for EISI of W$^{q+}$ ions with initial charge states $11\leq q \leq 18$. On the theory side, configuration-averaged distorted wave (CADW) calculations have been widely used for EISI calculations because of the efficiency of this method in generating large numbers of cross sections and plasma rate coefficients \cite{Loch2005,Pindzola2017,Pindzola2019a,Schury2020}. More elaborate level-resolved calculations have been carried out only for a limited number of ion species such as W$^{5+}$ \cite{Jonauskas2019a}, W$^{17+}$ \cite{Zhang2014}, W$^{25+}$ \cite{Kyniene2016}, W$^{26+}$ \cite{Kyniene2015}, and W$^{27+}$ \cite{Jonauskas2015}.

Recently, we have conducted a comprehensive comparison between subconfiguration-averaged distorted wave (SCADW) calculations (the relativistic variant of the CADW method) and level-to-level distorted wave (LLDW) calculations for EISI of W$^{14+}$ \cite{Jin2020}. It turned out that the SCADW and LLDW results were in good agreement with one another except for electron energies close to the ionisation threshold where important contributions by excitation autoionisation (EA) processes are neglected by the SCADW method. To remedy this deficiency of the otherwise efficient SCADW method we have proposed a hybrid approach, where only the EA cross sections in the vicinity of the ionisation threshold are calculated by the costly LLDW method. For W$^{14+}$ this hybrid cross section was in excellent agreement with the corresponding experimental result of Schury \etal \cite{Schury2020}.

In order to provide evidence for the wider applicability of this approach, we here present new hybrid SCADW+LLDW calculations for EISI of W$^{15+}$ and W$^{16+}$ ions and comparisons of the resulting cross sections with the corresponding experimental data of Schury \etal \cite{Schury2020}. In the calculations we also consider contributions by resonant-excitation--double-autoionisation (REDA) and by the possible presence of excited primary ions in the experiment.

\section{\label{method}Theoretical method}

Our theoretical approach, which is based on SCADW and LLDW methods as implemented in the Flexible Atomic Code (FAC) \cite{Gu2008}, has already been described extensively in our previous publication on EISI of W$^{14+}$ \cite{Jin2020}. Therefore, we here only provide details pertaining to the specific  W$^{15+}$ and W$^{16+}$ ion species of the present study. The ground configurations of these ions are [Kr]$4s^{2}4p^{6}4d^{10}4f^{11}5s^{2}$ and [Kr]$4s^{2}4p^{6}4d^{10}4f^{11}5s$, respectively \cite{Kramida2009}.

Our theoretical description of EISI includes direct ionisation (DI), EA, and REDA processes. A comprehensive review of these and other electron-ion interaction processes has been given by M\"uller \cite{Mueller2008a}. In the present calculations for W$^{15+}$ and W$^{16+}$, we account for DI of $5s$, $4f$, $4d$, $4p$ and $4s$ electrons. The EA process proceeds via two subsequent steps, where in a first step an inner-shell electron is excited by electron-impact such that a multiply excited state is formed which, in a second step, decays via the emission of an Auger electron. Here, we consider EA via the excitation of a $4s$, $4p$ or $4d$ electron to a higher subshell $nl$, i.e.,
\begin{eqnarray}
e&+&4s^{2}4p^{6}4d^{10}4f^{11}5s^x \nonumber \\
& &~\hspace*{1cm}\rightarrow\left\{
\begin{array}{l}
4s^{2}4p^{6}4d^{9\phantom{1}}4f^{11}5s^xnl\\
4s^{2}4p^{5}4d^{10}4f^{11}5s^xnl\\
4s\phantom{^2}4p^{6}4d^{10}4f^{11}5s^xnl
\end{array}
\right\}+e\label{eq:EA}
\end{eqnarray}
where $x=2$ for W$^{15+}$, $x=1$ for W$^{16+}$, $n_\mathrm{min}\leq n\leq n_\mathrm{max}$ and $0\leq l\leq l_\mathrm{max}$.  The maximum values $n_\mathrm{max}=25$ and $l_\mathrm{max}=8$  were chosen such that the EA cross sections are practically converged. The minimum value $n_\mathrm{min}$ is different for the individual EA channels. It is determined by the condition that the excited configurations or levels have to be in the ionisation continuum. For example, EA via the excitation of a $4f$ electron is energetically possible only for $n\geq n_\mathrm{min}=19$. Therefore it is assumed to be negligible, here. Moreover, it turns out that, for the present ionic systems, $n_\mathrm{min}$ also depends on the computational method as is discussed in detail below.

Also the REDA process can be viewed as a multi-step process with a resonant dielectronic capture (DC, time-inverse of the Auger process) as the first step which is followed by a cascade of two consecutive Auger processes. For W$^{15+}$ ($x=2$) and W$^{16+}$ ($x=1$) we consider the following DC channels in our calculations:
\begin{eqnarray}
e&+&4s^{2}4p^{6}4d^{10}4f^{11}5s^{x} \nonumber \\
& &~\hspace{1cm}\rightarrow \left\{
\begin{array}{l}
4s^{2}4p^{6}4d^{9\phantom{1}}4f^{11}5s^{x}n'l'nl\\
4s^{2}4p^{5}4d^{10}4f^{11}5s^{x}n'l'nl\\
4s\phantom{^2}4p^{6}4d^{10}4f^{11}5s^{x}n'l'nl
\end{array}
\right.,
\label{eq:REDA}
\end{eqnarray}
where $4\leq n'\leq 9$, $0\leq l'\leq 6$, $n'\leq n\leq 18$, and $0\leq l\leq 6$. As for EA, the maximum values of the above quantum numbers were chosen such that convergence is reached. In the subsequent Auger cascade, all required branching ratios for eventually arriving at the REDA cross sections were explicitly calculated. For the comparison with the experimental cross sections the theoretical REDA cross sections were convolved with a Gaussian to mimic the finite experimental energy spread of 2 eV full-width-at-half-maximum (FWHM).

In order to account for EISI of long-lived metastable levels that might have been present in the experiment \cite{Schury2020}, we also performed EISI calculations for the lowest excited configurations
[Kr]$4s^{2}4p^{6}4d^{10}4f^{12}5s$ and [Kr]$4s^{2}4p^{6}4d^{10}4f^{13}$ of W$^{15+}$ and [Kr]$4s^{2}4p^{6}4d^{10}4f^{12}$ and [Kr]$4s^{2}4p^{6}4d^{10}4f^{10}5s^2$ of W$^{16+}$ in the same manner as described above for the respective ground configurations.

\section{\label{sec:results}Results and discussions}

\subsection{SCADW calculations}\label{sec:SCADW}

\begin{table}
	\caption{\label{TABLE1}Degeneracies $g$ and excitation energies of the relevant subconfigurations of W$^{15+}$, W$^{16+}$, and W$^{17+}$. The energies in the one but last and last columns are relative to the lowest relativistic configurations of W$^{15+}$ and W$^{16+}$, respectively.}
\begin{indented}
	\item[]	
    \begin{tabular}{lrlrrr}
           \br
			Ion        & Index & Subconfiguration                     & $g$  & \multicolumn{2}{c}{Energy (eV)} \\
			\mr
			W$^{15+}$  & 0 & $4f_{5/2}^{6}4f_{7/2}^{5}5s_{1/2}^2$     & 56   & 0      &  \\
			           & 1 & $4f_{5/2}^{5}4f_{7/2}^{6}5s_{1/2}^2$     & 168  & 3.48   &  \\
		               & 2 & $4f_{5/2}^{4}4f_{7/2}^{7}5s_{1/2}^2$     & 120  & 5.69   &  \\
		               & 3 & $4f_{5/2}^{3}4f_{7/2}^{8}5s_{1/2}^2$     & 20   & 6.63   &  \\
		               & 4 & $4f_{5/2}^{6}4f_{7/2}^{6}5s_{1/2}$       & 56   & 7.55   &  \\
		               & 5 & $4f_{5/2}^{5}4f_{7/2}^{7}5s_{1/2}$       & 96   & 10.42  &  \\
		               & 6 & $4f_{5/2}^{4}4f_{7/2}^{8}5s_{1/2}$       & 30   & 12.05  &  \\
		               & 7 & $4f_{5/2}^{6}4f_{7/2}^{7}$               & 8    & 21.33  &  \\
		               & 8 & $4f_{5/2}^{5}4f_{7/2}^{8}$               & 6    & 23.62  &  \\[2ex]
			W$^{16+}$  & 9 & $4f_{5/2}^{6}4f_{7/2}^{5}5s_{1/2}$       & 112  & 360.05 & 0 \\
			           & 10& $4f_{5/2}^{6}4f_{7/2}^{4}5s^{2}_{1/2}$   & 70   & 361.94 & 1.89 \\
			           & 11& $4f_{5/2}^{5}4f_{7/2}^{6}5s_{1/2}$       & 336  & 363.56 & 3.51  \\
			           & 12& $4f_{5/2}^{6}4f_{7/2}^{6}$   & 28  & 364.52 & 4.74 \\
			           & 13& $4f_{5/2}^{4}4f_{7/2}^{7}5s_{1/2}$       & 240  & 365.80 & 5.75  \\
			           & 14& $4f_{5/2}^{5}4f_{7/2}^{5}5s^{2}_{1/2}$   & 336  & 366.08 & 6.03 \\
			           & 15& $4f_{5/2}^{3}4f_{7/2}^{8}5s_{1/2}$       & 40   & 366.75 & 6.70  \\
			           & 16& $4f_{5/2}^{5}4f_{7/2}^{7}$   & 48  & 367.69 & 7.64 \\
			           & 17& $4f_{5/2}^{4}4f_{7/2}^{6}5s^{2}_{1/2}$   & 420  & 368.93 & 8.88 \\
			           & 18& $4f_{5/2}^{4}4f_{7/2}^{8}$   & 15  & 369.33 & 9.28 \\
			           & 19& $4f_{5/2}^{3}4f_{7/2}^{7}5s^{2}_{1/2}$   & 160  & 370.47 & 10.42 \\
			           & 20& $4f_{5/2}^{2}4f_{7/2}^{8}5s^{2}_{1/2}$   & 15  & 370.71 & 10.66 \\	[2ex]
			W$^{17+}$  & 21& $4f_{5/2}^{6}4f_{7/2}^{5}$               & 56   & 745.50 & 385.45 \\
			           & 22& $4f_{5/2}^{5}4f_{7/2}^{6}$               & 168  & 749.05 & 389.00 \\
			           & 23& $4f_{5/2}^{4}4f_{7/2}^{7}$               & 120  & 751.30 & 391.25 \\
			           & 24& $4f_{5/2}^{3}4f_{7/2}^{8}$               & 20   & 752.57 & 392.52 \\
\br
		\end{tabular}
	\end{indented}
\end{table}

In a fully relativistic configuration-averaged calculation, the ground configurations of W$^{15+}$, W$^{16+}$, and W$^{17+}$ are split into subconfigurations. The energies of the here relevant subconfigurations are provided in table \ref{TABLE1}. The presently calculated ionisation energies of W$^{15+}$ and W$^{16+}$ are 360.05~eV and 385.45~eV, respectively. Both these energies are lower by 1.85~eV and 2.55~eV, respectively, than the corresponding values from the NIST Atomic Spectra Database (ASD) \cite{nist}.  This less than 1\% deviation is almost within the quoted uncertainties of the NIST values $361.9\pm1.5$ and $387.9\pm1.2$~eV.

\begin{figure}
\includegraphics[width=0.95\columnwidth]{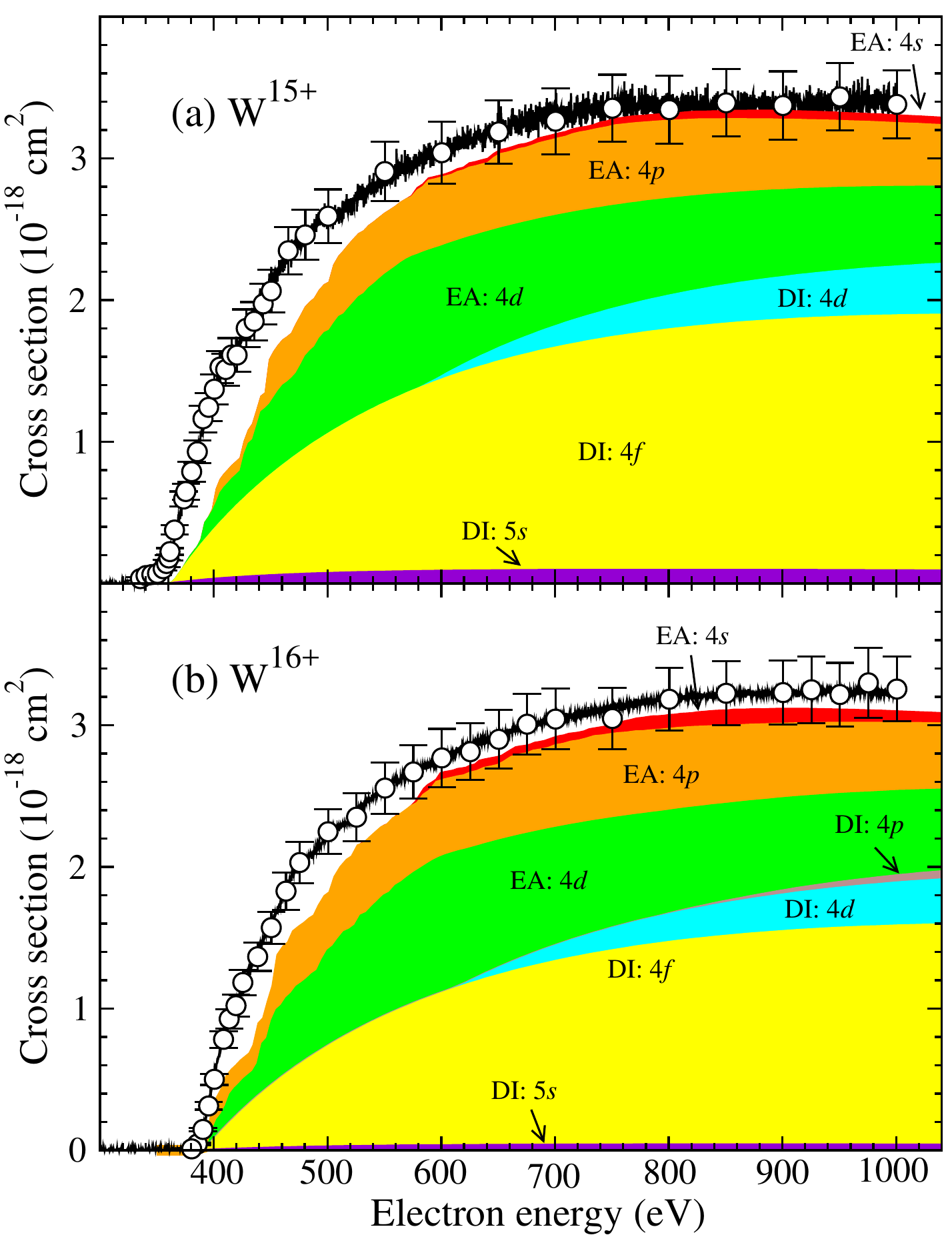}
\caption{\label{fig1} Comparison of the present SCADW cross sections (lines) for EISI of (a)  W$^{15+}$ and (b) W$^{16+}$ with the corresponding experimental results of Schury \etal \cite{Schury2020} (large open symbols: absolute cross-section data points with the error bars comprising statistical and systematic uncertainties; small filled symbols: scan measurements with the error bars comprising statistical uncertainties only). Contributions by different direct ionisation (DI) and excitation-autoionisation (EA) processes are represented by the labelled shaded curves.}
\end{figure}

Figure \ref{fig1} presents comparisons of our SCADW cross sections for EISI of W$^{15+}$ and W$^{16+}$ ions in their ground-configurations with the corresponding experimental data of Schury \etal \cite{Schury2020}. In the SCADW calculations we assumed a statistical population of the subconfigurations. For both ions, the correspondingly averaged cross sections agree well with the experimental data for electron energies beyond 600 eV. At lower energies closer to the ionisation thresholds, the SCADW cross sections significantly underestimate the experimental cross sections. This corresponds to our earlier findings for EISI of W$^{14+}$ \cite{Jin2020}. Our calculations show that the dominating ionisation processes for both ions are DI of a $4f$ electron and EA involving the excitation of a $4p$ or a $4d$ electron. The thresholds for DI of a $4p$ or a $4s$ electron occur beyond the thresholds for double ionisation. Therefore, these ionisation channels contribute almost negligibly to the total cross sections for single ionisation. Further minor contributions stem from DI of an outer $5s$ electron and from EA of a $4s$ electron.

\begin{figure}
\includegraphics[width=0.95\columnwidth]{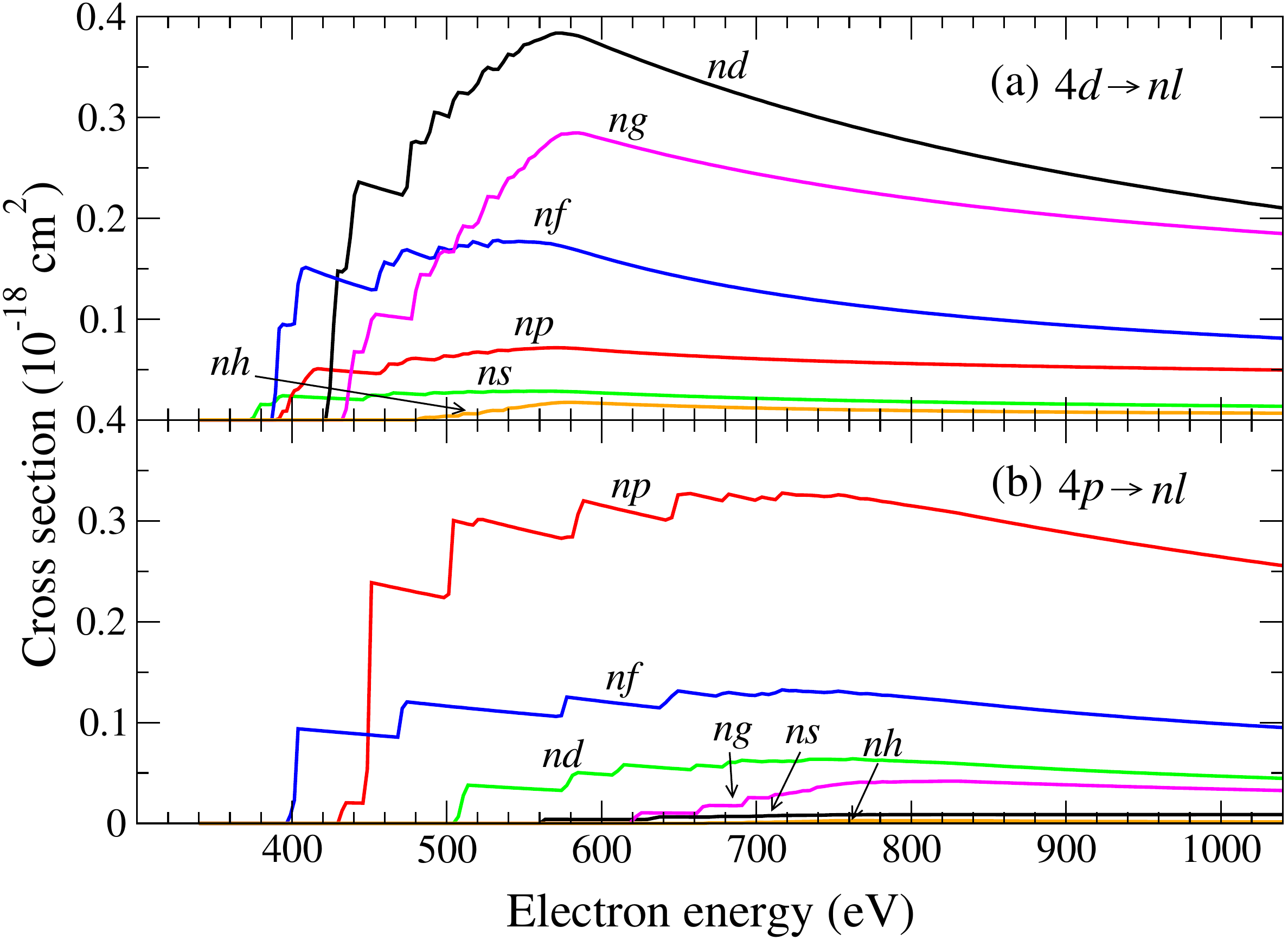}
\caption{\label{fig2} SCADW EA contributions to the total cross section for EISI of W$^{15+}$: (a) $4d\rightarrow nl$ and (b) $4p\rightarrow nl$ summed over $n$ up to $n=25$.}
\end{figure}

There are two major causes for the  disagreement between SCDAW theory and experiment in the near threshold region, i.e., i) a possible presence of metastable ions in the ion beams of the experiment and ii) deficiencies in the theoretical treatment of EA. The presence of metastable primary ions reveals itself by nonzero experimental cross sections below the threshold for ionisation of ions in the ground subconfiguration.  We will discuss this issue in more detail below (section \ref{sec:hybrid}). First, we focus on the EA process which is particulary important close to the ionisation threshold (figure~\ref{fig1}).

Figure \ref{fig2} shows the important $4d\rightarrow nl$ and $4p\rightarrow nl$ EA contributions to EISI of W$^{15+}$ as calculated by the SCADW method. For a given angular momentum $l$, the $nl$ channels are summed over the principal quantum number $n$ up to $n=25$. In the SCADW calculations these channels all open up only beyond the ionisation threshold. The strongest $4d\rightarrow nd$ EA channels open at 421~eV. This energy corresponds to the $4d\rightarrow 6d$ excitation energy. The SCADW $4d\rightarrow 5d$ excitation energies appear to be below the EISI threshold at 360.05~eV for all 16 subconfigurations of the $4d^{9}4f^{11}5s^{2}5d$ configuration which span the energy range $319.97-342.89$~eV. Thus, the $4d\rightarrow 5d$ EA channel is entirely missing in the SCADW calculations for W$^{15+}$ and the same is found here also for W$^{16+}$. This situation is different from our earlier findings for W$^{14+}$ where some of the excited $4d^{9}4f^{12}5s^{2}5d$ subconfigurations were below and some above the W$^{14+}$ single-ionisation threshold \cite{Jin2020}.

\subsection{LLDW calculations of $4d\to5d$ EA}\label{sec:LLDW}

Our LLDW calculations reveal that the $4d\rightarrow 5d$ EA channel cannot be entirely neglected for W$^{15+}$ and W$^{16+}$, even though it does not contribute to EISI of these ions in the SCADW approach. In the more detailed level-to-level approach, the $4p^{9}4f^{11}5s^{2}5d$ configuration of W$^{15+}$ splits into 3565 levels with excitation energies ranging from 307.57 to 383.49~eV. Although most of these levels are still below the ionisation threshold, 440 are above, i.e.,  in the LLDW calculation there are contributions by the $4d\rightarrow 5d$ EA channel to the EISI cross section of W$^{15+}$.

\begin{figure}
\includegraphics[width=0.95\columnwidth]{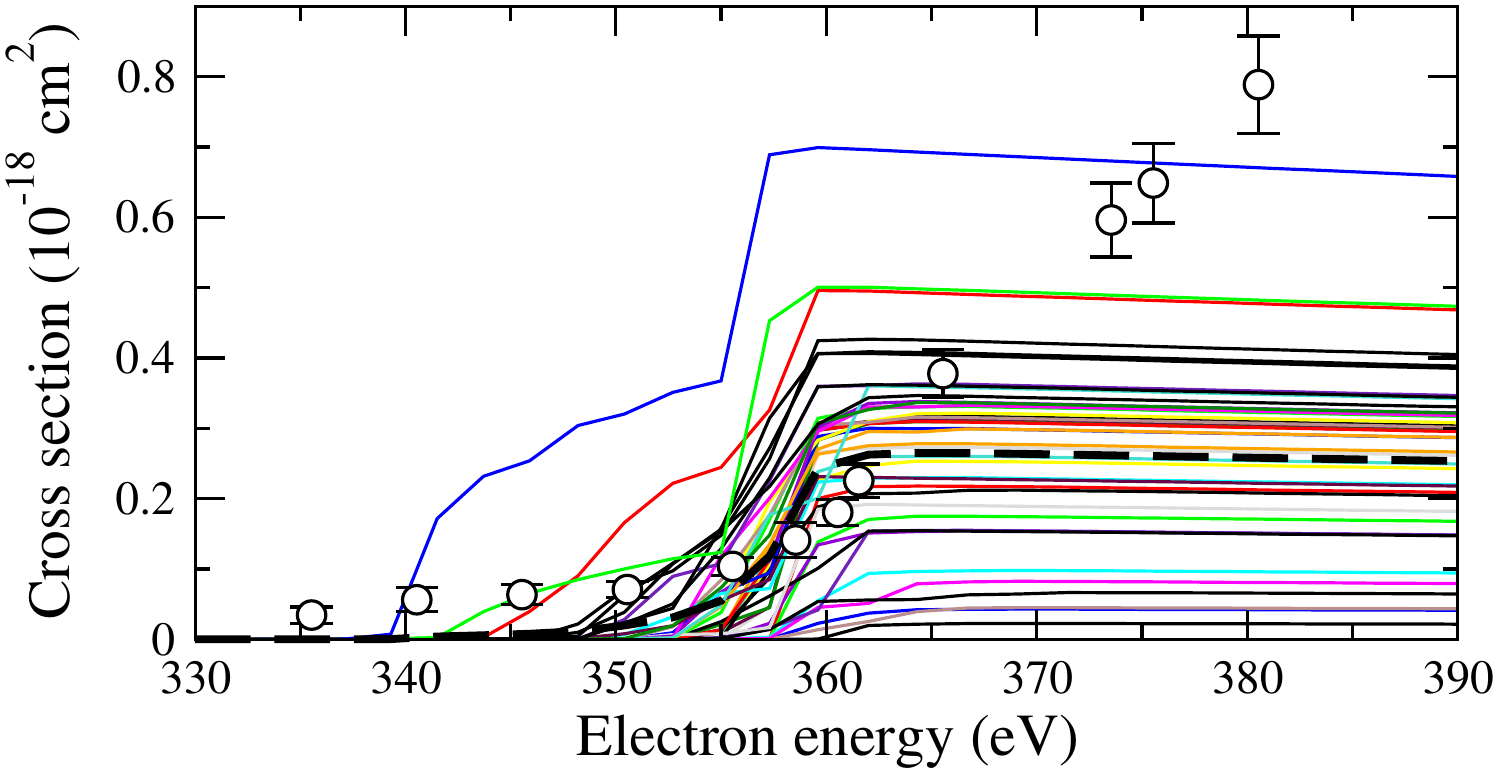}
\caption{\label{fig3} LLDW $4d\rightarrow 5d$ EA cross sections for the 41 levels of the W$^{15+}$ ground configuration (colored solid lines) The thick black dashed line is the statistically weighted average of the individual contributions. The open symbols are the absolute cross-section data points for EISI of W$^{15+}$ measured by Schury \etal \cite{Schury2020} with the error bars comprising statistical and systematic uncertainties.}
\end{figure}

In the LLDW approach, also the ground configuration of the primary ion splits into several fine-structure levels. For the $4d^{10}4f^{11}5s^{2}$  configuration of W$^{15+}$ this results in 41 levels with energies of up to 23.14~eV above the ground-level energy. This splitting is much larger than that of the subconfigurations in the SCADW calculation (table \ref{TABLE1}).  The thin colored lines in figure \ref{fig3} represent the individual $4d\rightarrow 5d$ EA cross sections for each of the 41 levels of the W$^{15+}$ ground configuration.

Obviously, the individual levels contribute differently to the averaged EA cross section (thick black dashed line in figure \ref{fig3}). For the averaging we assumed a statistical population of all levels, since their lifetimes are much longer than the microsecond flight time of the ions in the experiment of Schury \etal \cite{Schury2020}. This is because the transitions from the excited levels to the ground level are dipole forbidden. For example, for the highest level, i.e., the $[( 4f_{5/2}^{4})_{0}(4f_{7/2}^{7})_{7/2}5s_{1/2}^{2}]_{7/2}$ level, we calculate a lifetime of 1.2~ms taking electric quadrupole (E2) and magnetic dipole (M1) transitions into account. This level has the largest $4d\to5d$ EA cross section (thin blue line in figure \ref{fig3}). Clearly, the level-averaged cross section is very sensitive to its population.

\begin{figure}
\includegraphics[width=0.95\columnwidth]{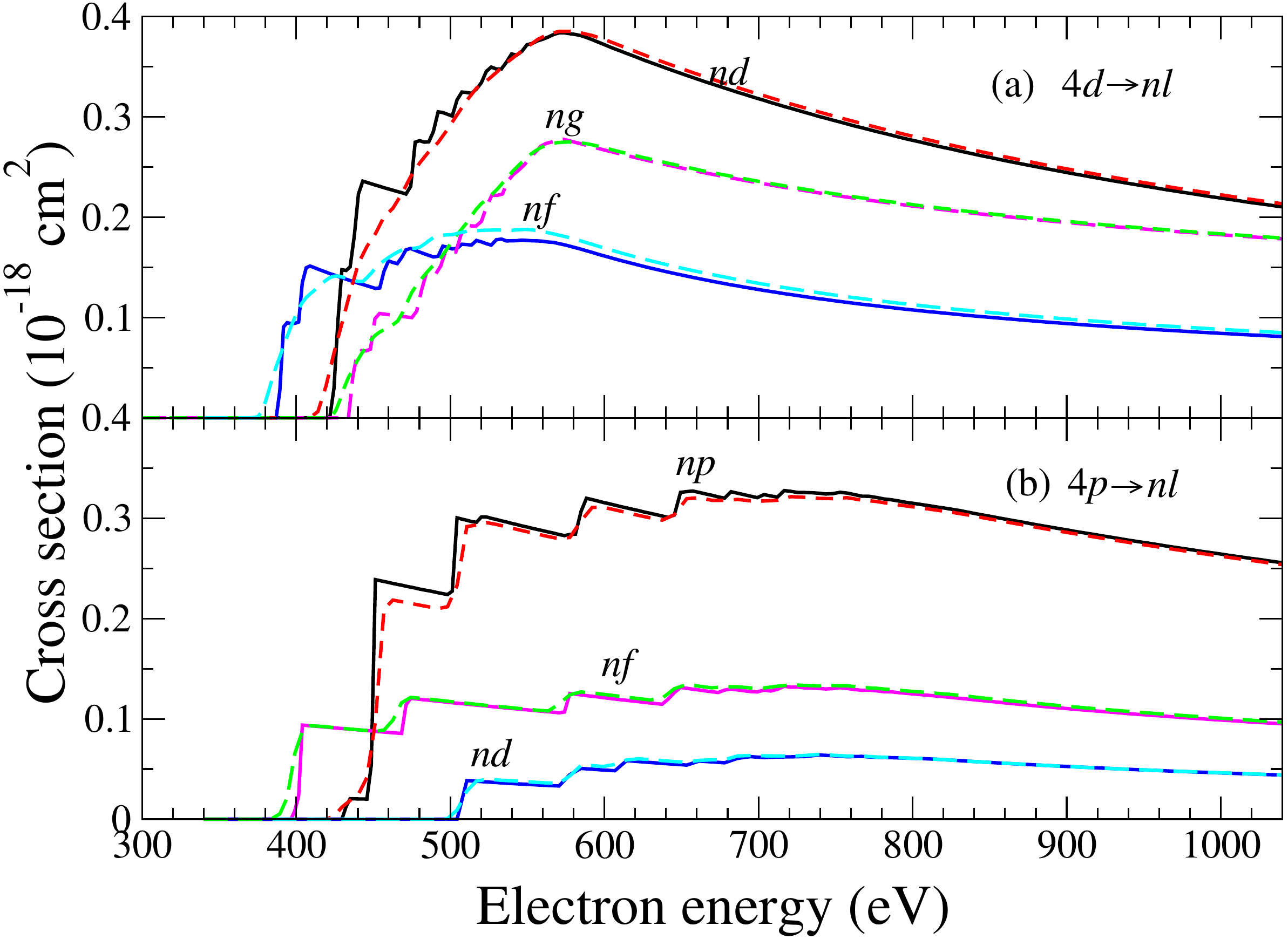}
\caption{\label{fig4} Comparison between the LLDW (dashed lines) and SCADW (full lines) results for the $4d\rightarrow nl$ and $4p\rightarrow nl$ EA channels (except for the $4d\to5d$ channnel) of the W$^{15+}$ ground configuration with $n\leq 13$.}
\end{figure}

Irrespective of the detailed level populations, the comparison between experiment and the LLDW $4d\to5d$ EA cross section in figure \ref{fig3} strongly suggests that  the neglect of this EA channel in the SCADW calculation is largely responsible for the  disagreement between SCADW and experimental near-threshold cross section for EISI of W$^{15+}$. Figure \ref{fig4} shows, that there are only minor differences between the SCADW and LLDW calculations for the remaining EA channels.

\begin{figure}
\includegraphics[width=0.95\columnwidth]{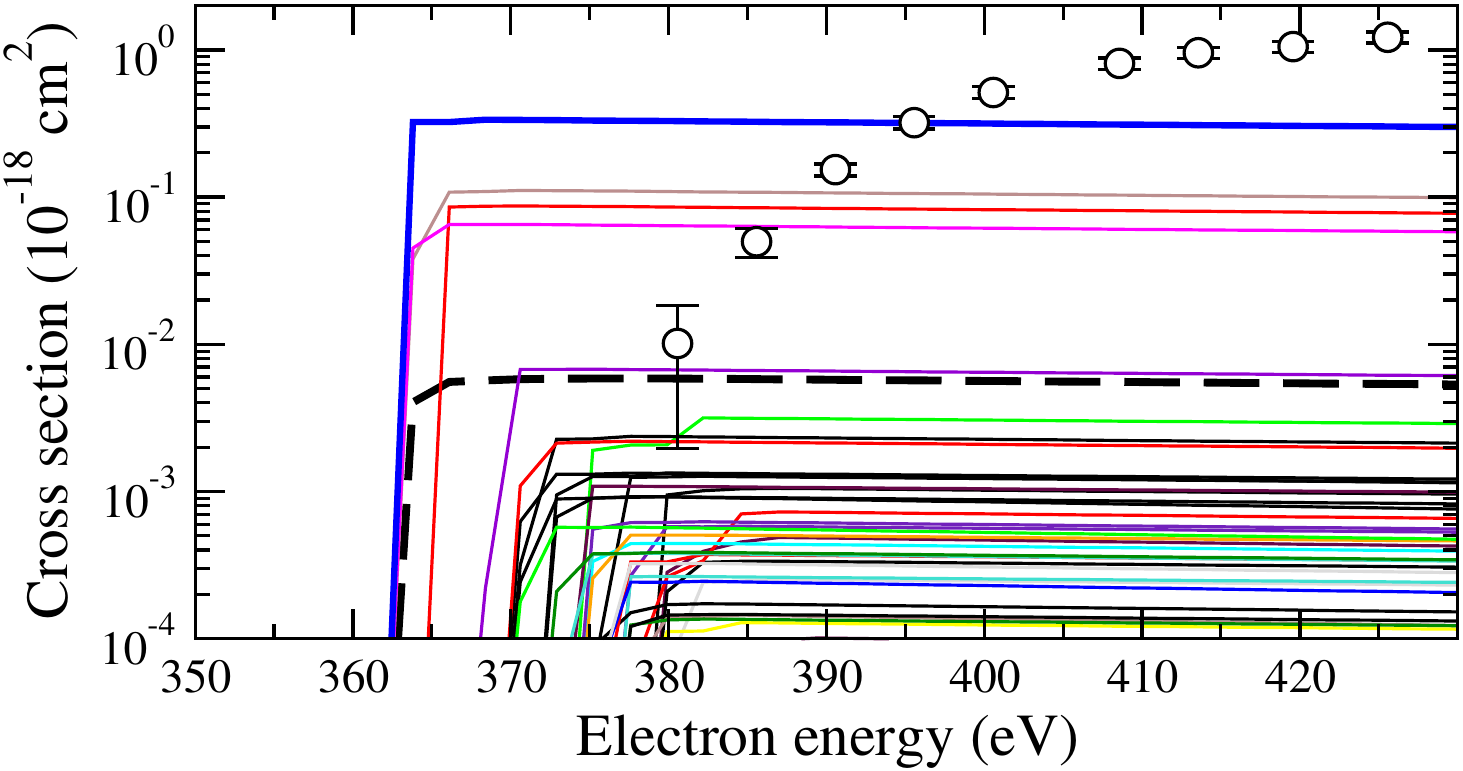}
\caption{\label{fig5} LLDW $4d\rightarrow 5d$ EA cross sections for the 82 levels of the W$^{16+}$ ground configuration (colored solid lines). The cross-section scale is logarithmic to visualize also the smaller cross sections. The thick black dashed line is the statistically weighted average of the individual contributions. The open symbols are the absolute cross-section data points for EISI of W$^{16+}$ measured by Schury \etal \cite{Schury2020} with the error bars comprising statistical and systematic uncertainties.}
\end{figure}

The $4d^{10}4f^{11}5s$ ground configuration of W$^{16+}$ splits into 82 levels. Their individual contributions to the $4d\to5d$ EA cross section are displayed in figure \ref{fig5}. For W$^{16+}$ the statistically weighted average of the individual cross sections (thick black dashed line in figure~\ref{fig5}) is dominated by only a few levels with the strongest contribution of up to $2\cdot10^{-19}$~cm$^2$ stemming from  the $[(4f_{5/2}^{4})_{0}(4f_{7/2}^{7})_{7/2}5s_{1/2}]_{3}$ level (thick blue line in figure~\ref{fig5}). In comparison, there are 77 channels whose cross section maxima are below $3\cdot10^{-21}$~cm$^2$. As a result, the statistically averaged $4d\to5d$ EA cross section is an order of magnitude smaller for W$^{16+}$ as compared to W$^{15+}$.

\subsection{Hybrid SCAWD+LLDW cross sections}\label{sec:hybrid}

\begin{figure}
\includegraphics[width=0.95\columnwidth]{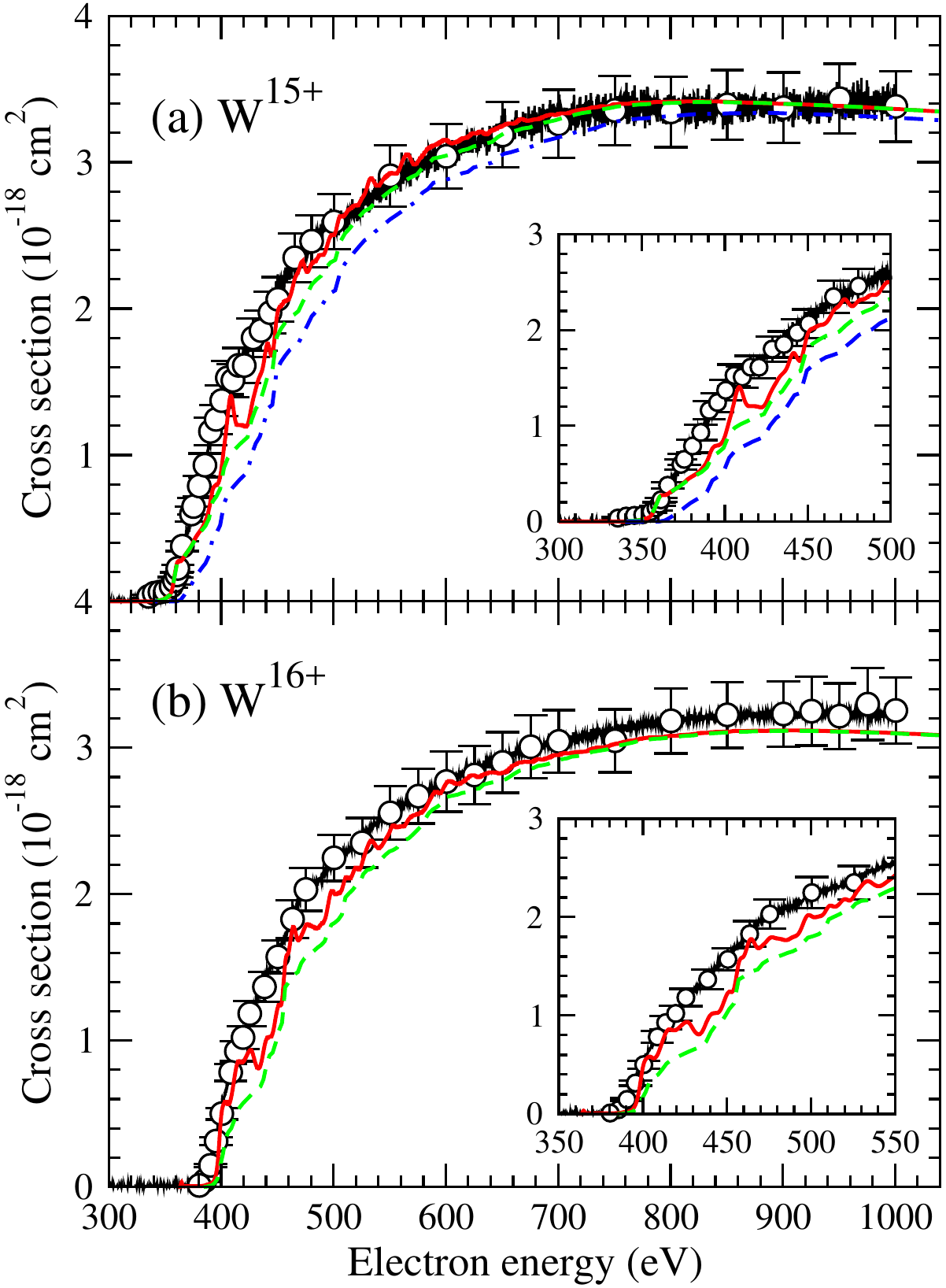}
\caption{\label{fig6} Total cross sections for EISI of (a) W$^{15+}$ and (b) W$^{16+}$. The open and closed symbols are the experimental data measured by Schury \etal \cite{Schury2020} (see figure \ref{fig1}).  The green dashed lines represent the hybrid cross sections (SCADW + LLDW for $4d\rightarrow 5d$ EA).  The red full lines represent the hybrid cross sections including REDA contributions in addition. The blue dash-dotted line in panel (a) is the SCADW result for W$^{15+}$ from figure~\ref{fig1}. The corresponding curve for W$^{16+}$ is not shown since it is almost indistinguishable from the hybrid cross section (green dashed line).  The insets magnify the near threshold regions.}
\end{figure}

Figure \ref{fig6} shows our hybrid cross sections for EISI of W$^{15+}$ and W$^{16+}$. These were obtained by simply adding the statistically averaged LLDW cross sections for $4d\to 5d$ EA from section~\ref{sec:LLDW} to the SCADW results from section~\ref{sec:SCADW}. As expected, the agreement between the calculated and experimental cross sections is improved when the $4d\to 5d$ EA channel is taken into account. The improvement is more significant for W$^{15+}$ than for W$^{16+}$ where the  $4d\to 5d$ EA cross section is almost negligible. For both ions, the agreement becomes even better if REDA is taken into account in addition.

\begin{figure}
\includegraphics[width=0.95\columnwidth]{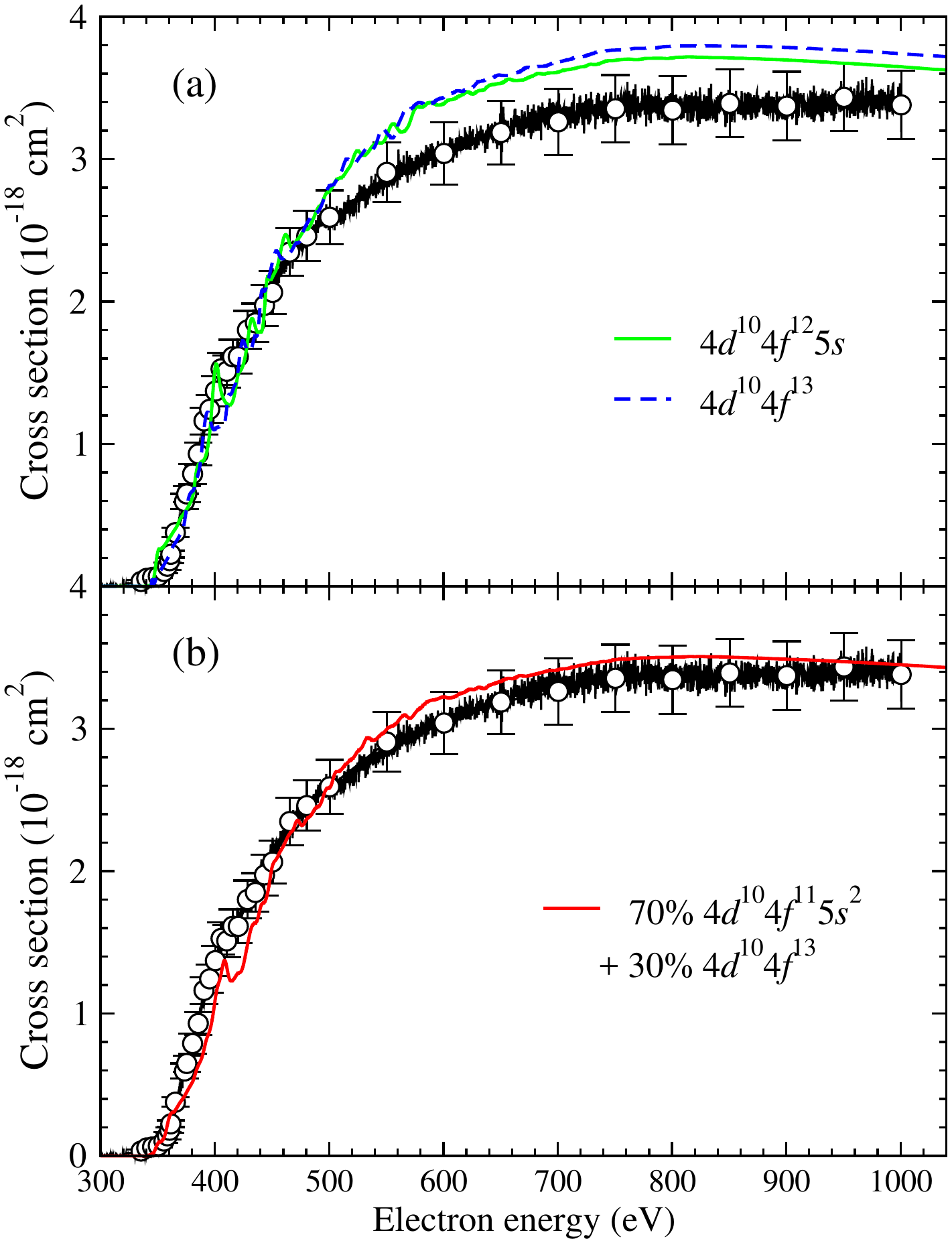}
\caption{\label{fig7} Hybrid cross sections for EISI of the two lowest excited configurations of W$^{15+}$, i.e., the  $4d^{10}4f^{12}5s$ configuration (green solid line) and the $4d^{10}4f^{13}$ (blue dashed line) in comparison with the experimental data of Schury \etal \cite{Schury2020} (see figure~\ref{fig1}). The red full line in panel (b) is the cross section for a mixture of 70\% $4d^{10}4f^{11}5s^2$ ground configuration and 30\% $4d^{10}4f^{13}$ excited configuration.}
\end{figure}

\begin{figure}
\includegraphics[width=0.95\columnwidth]{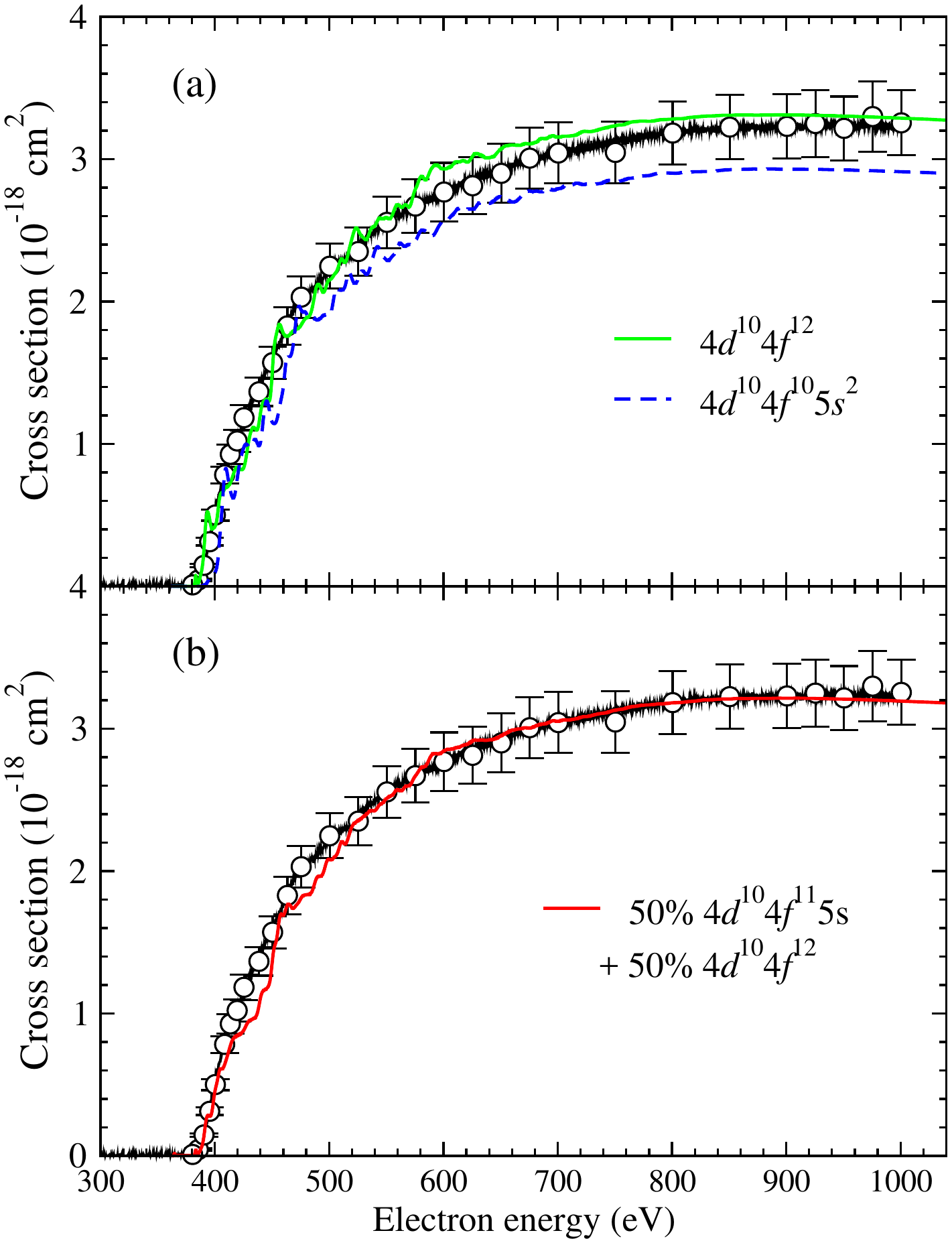}
\caption{\label{fig8} Hybrid cross sections for EISI of the two lowest excited configurations of W$^{16+}$, i.e.,  the $4d^{10}4f^{12}$ configuration (green solid line) and the $4d^{10}4f^{10}5s^{2}$ configuration (blue dashed line) in comparison with the experimental data of Schury \etal \cite{Schury2020} (see figure~\ref{fig1}). The red full line in panel (b) is the cross section for a mixture of  50\%  $4d^{10}4f^{11}5s$ ground configuration and 50\% $4d^{10}4f^{12}$ excited configuration.}
\end{figure}

The calculations so far were confined to the W$^{15+}$ and W$^{16+}$ ground configurations. However, in experiments with ion beams, depending on the ion production process, certain fractions of ions can be in excited metastable configurations. Since the ionisation cross sections of these ions differ from the cross sections for ions in the ground configuration, the presence of metastable ions potentially influences the measured cross sections. Usually, the metastable fractions of ions are not known and must be inferred from detailed comparisons with theoretical calculations as shown recently, e.g., for W$^{5+}$ \cite{Jonauskas2019a}.

As discussed in some more detail by Schury \etal \cite{Schury2020} the influence of metastable ions on the measured EISI cross section for W$^{15+}$ and W$^{16+}$ is expected to be rather small. Nevertheless, we here performed SCADW+LLDW hybrid calculations (including REDA) of EISI cross sections for W$^{15+}$ and W$^{16+}$ ions in initially excited configurations in addition to our calculations for ground-configuration ions. For both ions, the additional configurations considered are the two lowest excited configurations, i.e., the $4d^{10}4f^{12}5s$ and $4d^{10}4f^{13}$  configurations for W$^{15+}$ and the $4d^{10}4f^{12}$ and  $4d^{10}4f^{10}5s^{2}$ configurations of W$^{16+}$. The corresponding subconfigurations are listed in table~\ref{TABLE1}.

Figure~\ref{fig7}(a) shows the hybrid cross sections (including REDA) for the two lowest excited configurations of W$^{15+}$.  Near to the ionisation threshold these cross sections agree very well with the experimental data, however, at energies above about 500~eV they are significantly larger. An unambiguous determination of the excited beam fractions in the experiment on the basis of our calculations is not possible, although the overall agreement between experiment and theory can be improved if some (arbitrary) assumptions about the excited fractions are made as shown in figure \ref{fig7}(b). This is also the case for W$^{16+}$ (figure~\ref{fig8}).

\section{Conclusions}

In summary, we have performed detailed calculations of the total cross sections for EISI of W$^{15+}$ and W$^{16+}$ ions by using a hybrid SCADW+LLDW approach. For both ions, costly LLDW calculations were performed for the EA processes involving $4d\to5d$ excitations. In the SCADW approach these are entirely neglected because all SCADW $4d\to5d$ excitation energies are below the single-ionisation threshold. The hybrid SCADW+LLDW cross sections agree very well with the experimental data of Schury \etal \cite{Schury2020} if REDA processes are included in addition to DI and EA.  Our calculations for ions in initially excited configurations suggest that such ions only have a minor influence on the experimental cross sections for EISI of W$^{15+}$ and W$^{16+}$. As already found previously for W$^{14+}$~\cite{Jin2020}, the hybrid SCADW+LLDW approach for calculating total EISI cross sections here proves again to be capable of delivering reliable results for complex ions at moderate expense.

\section*{Acknowledgments}

This work is supported by National Science Foundation of China (grant no. 11374365). Financial support by the German Federal Ministry of Education and Research (BMBF) within the \lq\lq{}Verbundforschung\rq\rq\ funding scheme (grant no.\ 05P19RGFA1) is gratefully acknowledged. B.E. is financially supported by a grant provided within the frame of the formal cooperation between the GSI Helmholtzzentrum f\"ur Schwerionenforschung (Darmstadt, Germany) and the Justus-Liebig-Universit\"at Gie{\ss}en.

\section*{References}


\providecommand{\newblock}{}

\end{document}